\documentclass[12 pt, amsfonts,amsmath, amssymb,color]{article}
\usepackage{authblk}
\def\baselinestretch{1.0}
\usepackage{amsmath,amssymb,amsfonts,latexsym,float,graphics,epsfig}
\usepackage{subfig}
\usepackage{verbatim}
\usepackage{relsize}
\usepackage{hyperref}
\usepackage{graphicx}
\usepackage{bm}
\usepackage{epstopdf}
\usepackage{slashed}
\usepackage{color}
\usepackage{tikz-cd}
\usepackage[utf8]{inputenc}
\usepackage[T1]{fontenc}

\def\be{\begin{equation}}
\def\ee{\end{equation}}
\def\bea{\begin{eqnarray}}
\def\eea{\end{eqnarray}}

\begin{document}

\renewcommand\theequation{\arabic{section}.\arabic{equation}}
\catcode`@=11 \@addtoreset{equation}{section}
\newtheorem{axiom}{Definition}[section]
\newtheorem{theorem}{Theorem}[section]
\newtheorem{axiom2}{Example}[section]
\newtheorem{lem}{Lemma}[section]
\newtheorem{prop}{Proposition}[section]
\newtheorem{cor}{Corollary}[section]

\newcommand{\ben}{\begin{equation*}}
\newcommand{\een}{\end{equation*}}

\let\endtitlepage\relax

\begin{titlepage}
\begin{center}
\renewcommand{\baselinestretch}{1.5}  

\vspace*{-0.5cm}

{\Large \bf{Statistical ensembles and logarithmic}} \\
 {\Large \bf{corrections to black hole entropy}}

\vspace{9mm}
\renewcommand{\baselinestretch}{1}  

\centerline{\large{Aritra Ghosh}\footnote{ag34@iitbbs.ac.in}}

\vspace{5mm}
\normalsize
\textit{School of Basic Sciences, Indian Institute of Technology Bhubaneswar,}\\
\textit{Argul, Jatni, Khurda, Odisha 752050, India}\\
\vspace{5mm}

\begin{abstract}
In this paper, we consider general statistical ensembles and compute logarithmic corrections to the microcanonical entropy resulting due to thermodynamic fluctuations which are controlled by the boundary conditions, i.e. due to choice of ensemble. The framework is applied to the case of non-extremal black holes to give certain logarithmic corrections to the Bekenstein-Hawking entropy. We argue that within the framework of black hole chemistry, where the cosmological constant is identified with bulk pressure, the isoenthalpic-isobaric entropy rather than microcanonical entropy carries a more natural and consistent thermodynamic interpretation as black hole entropy. Logarithmic corrections to both microcanonical and isoenthalpic-isobaric entropies of black holes are computed, and we show that the latter set of corrections in black hole chemistry are of the same form as corrections to the microcanonical entropy in theories where the cosmological constant is not interpreted as a thermodynamic pressure. Finally, we compute logarithmic corrections to entropy in the framework of holographic black hole chemistry. We emphasize upon the choice of statistical ensemble, both in the bulk and on the boundary, in order to have a consistent comparison between them. The corrections studied in this paper are distinct from those obtained from Euclidean quantum gravity.
\end{abstract}
\end{center}
\vspace*{0cm}


\end{titlepage}
\vspace*{0cm}
\clearpage

\tableofcontents

\clearpage
\section{Introduction}
\pagenumbering{arabic} 
Black holes are intriguing objects in both classical and quantum theories of gravity. Semi-classically, black holes are associated with the notions of entropy and temperature \cite{Bekenstein:1973ur,Bekenstein:1974ax,Hawking:1974sw,Hawking:1976de}, thereby leading to a thermodynamic interpretation which is consistent with the laws of thermodynamics. A test for any quantum theory of gravity is the successful computation of the black hole entropy via microscopic counting of states. In this regard, the connection between the macroscopic and microscopic formulations is based on Boltzmann's celebrated formula\footnote{We will set \(k_B = \hbar = c = 1\) in all subsequent discussions.}: 
\begin{equation}\label{Boltzmann}
S = \ln \Omega,
\end{equation}
where \(\Omega\) is the number of accessible microstates at equilibrium, each of which have equal probability. Over the past three decades, it has been possible to compute \(\Omega\) microscopically for various cases of extremal black holes in string theory (see for example \cite{Vafa}-\cite{SenLec} for some famous works) and the results match with the macroscopic Bekenstein-Hawking entropy \(S_{\rm BH}\). Even at non-zero temperature, there are some instances of successful matching of macroscopic and microscopic results, such as in the case of the Ba\~nados-Teitelboim-Zanelli (BTZ) black hole, where the Cardy's formula for the entropy of a \((1+1)\)-dimensional conformal field theory (CFT) \cite{Cardy} reduces to Bekenstein-Hawking entropy for large conformal dimension (via the Brown-Henneaux formalism \cite{BroHen}). Thus, one may regard the CFT states with large conformal dimension to describe the microstates of the BTZ black hole \cite{BTZStrom,BTZ}, an identification which follows from the fact that the asymptotic symmetry algebra associated with three-dimensional gravity gives rise to two copies of the Virasoro algebra describing the holomorphic and anti-holomorphic sectors of the CFT \cite{BroHen}. \\

Apart from computations of the entropy, there have been developments towards computing perturbative corrections to the black hole entropy in a variety of settings \cite{9407001}-\cite{ACSHol}. In general, such corrections take the form \cite{0406044}:
\begin{equation}\label{loggen}
S_{\rm corrected} = S_{\rm BH} - k \ln S_{\rm BH} + \cdots,
\end{equation} where \(k\) is some real constant depending upon the details of the theory. Loop quantum gravity (LQG) treatments have also lead to corrections of the logarithmic kind \cite{0002040,0401070} (see also \cite{9801080}-\cite{0905.3168}). In \cite{Carlip}, logarithmic corrections to the entropy of BTZ black holes were computed by considering such corrections to the Cardy's formula (see also \cite{0104010}). Perhaps one of the most widely used frameworks for computing such corrections is Euclidean quantum gravity, where one-loop corrections to the Euclidean action computed using the heat kernel expansion \cite{HKL} give rise to logarithmic corrections to black hole entropy (see for example \cite{Shamik,Bhattacharyya}). Such one-loop corrections include contributions from all the fields in the theory, and of course the graviton loop, which becomes the only contribution in a pure gravity theory. It is believed that higher loops do not lead to logarithmic corrections to the entropy because infrared divergences become `softer' at higher loops in gravity theories \cite{Sen:2012dw}. \\

Apart from these corrections, there are logarithmic corrections arising due to thermodynamic fluctuations, which are controlled by the boundary conditions in a statistical mechanics sense. For instance, if a system is kept in contact with a heat bath at temperature \(T_0\), then even at equilibrium, there can be small fluctuations in the temperature \(T\) of the system (about \(T_0\)). Naturally, such fluctuations lead to a larger number of accessible microstates than those that would be allowed if the temperature were to remain fixed at \(T = T_0\). The cumulative effect of these additional microstates manifests as logarithmic corrections to the entropy, even for black holes. To understand how such corrections arise, let us consider writing the number of microstates as [see subsection-(\ref{cansec}) for details] (see also \cite{Braden})
\begin{equation}
\Omega(E) \sim \int e^{\beta (F-E)} d\beta = \int e^{S(\beta)} d\beta,
\end{equation} where \(\beta = 1/T\), \(F\) is the free energy, \(E\) is the energy, and in the second equality, we used \(F = E - TS\). Thus, if \(\beta = \beta_0\) is the equilibrium value of inverse temperature, i.e. that corresponding to the saddle point, one can write 
\begin{equation}
\Omega(E) \approx e^{S_0},
\end{equation} where \(S_0 = S(\beta_0)\) and this gives Eq. (\ref{Boltzmann}). This is the `zero-loop' contribution to the density of states \cite{Braden}. However, if one were to consider fluctuations in the inverse temperature, however small they are, one needs to expand the function \(S(\beta)\) about \(\beta = \beta_0\) and integrate over the next non-zero order to obtain the first-order corrected density of states. This always gives the generic structure \(\Omega(E) \approx e^{S_0}/\sqrt{D}\), where \(D\) can be interpreted as the `one-loop determinant' in the statistical mechanics sense\footnote{We use this terminology following \cite{Braden}.}. This leads to logarithmic corrections to the microcanonical entropy and such corrections have been reported in various works in the context of black hole thermodynamics \cite{Pourhassan:2017qxi,parthasarathi,Mukherji:2002de,Chatterjee,Sarkar,open,CGQ,ACSHol} (see also \cite{Braden}). It should be noted that these corrections which stem from statistical thermodynamics arguments are distinct from those obtained as one-loop corrections in Euclidean quantum gravity. These thermodynamic corrections shall continue to be present even if the gravity theory does not lead to any corrections to the Euclidean partition function. However, we should keep in mind that the thermodynamic corrections are valid only at non-zero temperatures (more pronounced at high temperatures), and away from the critical point(s), where fluctuations are large resulting in the `higher-loop' contributions being significant. \\

The aim of the present paper is to discuss a coherent formalism describing such `one-loop corrections' to black hole entropy in a variety of ensembles, i.e. boundary conditions. We consider various black holes and obtain such corrections in a variety of settings. Some of these corrections are new while some of them were reported earlier but we include them anyway to make the presentation self-contained. We particularly emphasize upon the black hole chemistry framework, where the cosmological constant is treated as a thermodynamic variable \cite{Kastor:2009wy,Cvetic:2010jb,Dolan:2010ha,Dolan:2011xt,Dolan119,Kubiznak:2012wp,45} (see also \cite{Henneaux:1984ji,Teitelboim:1985dp,Henneaux:1989zc,Caldarelli:1999xj} for some older works), for which some studies on logarithmic corrections have been reported recently \cite{open,CGQ,ACSHol}. The isoenthalpic-isobaric ensemble is introduced to describe black holes in this framework, which was not considered earlier. Subsequently, we obtain corrections to the microcanonical as well as isoenthalpic-isobaric entropies of black holes. Entropy corrections are also discussed in the context of holographic black hole chemistry. \\

The paper is organized as follows. In the next section, we describe some generic statistical ensembles and discuss the one-loop corrections to entropy in ensembles with one or two fluctuating coordinates. We present the isoenthalpic-isobaric entropy and obtain corrections to it in the one-loop order for a general thermodynamic system. Then, in section-(\ref{BHsec}), we consider some selected examples of black holes and compute logarithmic corrections to the microcanonical entropy in one or two fluctuating variables which include energy, electric charge, and angular momentum. Subsequently, we study black holes in the framework of black hole chemistry in section-(\ref{ExtSec}) and analyze corrections to the microcanonical entropy as well as the isoenthalpic-isobaric entropy. Finally, we explore logarithmic entropy corrections in holographic black hole chemistry in section-(\ref{holsec}). We end with some discussion in section-(\ref{Disc}). Two appendices-(\ref{appA}) and (\ref{Q2}) are included as supplementary material. 

\section{Statistical ensembles}
In this section, we describe a few statistical ensembles and discuss how to compute logarithmic corrections to entropy in such settings. The framework shall be applied to black holes in sections-(\ref{BHsec}) and (\ref{ExtSec}). We shall be quite brief and the reader is referred to \cite{Pourhassan:2017qxi,parthasarathi,Mukherji:2002de,Chatterjee,Sarkar,open,CGQ,ACSHol}, where similar calculations for obtaining logarithmic corrections to black hole entropy have been discussed. We begin by discussing the canonical ensemble. 

\subsection{Canonical ensemble}\label{cansec}
Let us consider the familiar canonical ensemble where the system is in contact with a thermostat (bath) at temperature \(T (= 1/ \beta)\) and there are some extensive quantities \(X^1, X^2, \cdots\) which are fixed by the walls of the container. Therefore, the system attains thermodynamic equilibrium with fixed parameters \((T,X^i)\), where the system can exchange energy \(E\) with the thermostat to adjust its temperature to become equal to that of the latter. If one imposes the natural condition that the energy of the bath is very large as compared to that of the system, then the probability of finding the system with energy \(E_k\) is \cite{callen} \(P_k = C e^{-\beta E_k}\), where \(C\) is a normalizing factor which is determined from the condition \(\sum_k P_k = 1\), giving
\begin{equation}
C^{-1} = \sum_k e^{-\beta E_k}.
\end{equation} The quantity \(C^{-1}\) is the canonical partition function \(Z\) and it depends on the variables \((T,X^i)\) describing the equilibrium state of the system. In the continuum limit, one has
\begin{equation}
\sum_k \rightarrow \int_0^\infty \Omega(E) dE,
\end{equation} where \(\Omega(E)\) is the density of states, and therefore the partition function is defined as
\begin{equation}\label{CanZ}
Z = \int_0^\infty \Omega(E) e^{-\beta E} dE.
\end{equation}
It should be noted that in addition to \(E\), the density of states depends upon \(X^i\), i.e. \(\Omega = \Omega(E,X^i)\). We omit writing out this explicit dependence for the sake of brevity. In a typical statistical mechanics textbook problem, one is aware of the energy spectrum and is able to determine the number of microstates in the energy interval \(E\) to \(E + dE\). Then, finding out the partition function amounts to computing the integral given in Eq. (\ref{CanZ}). On the other hand, there are systems such as black holes, where the microscopic density of states \(\Omega\) is not known in general. However, macroscopic thermodynamic functions such as mass, entropy, angular momentum and electric charge together with the relations between them are known from the gravity theory. In what follows, we will describe a procedure to compute logarithmic corrections to the entropy of such systems from the knowledge of macroscopic response functions \cite{parthasarathi,callen}. \\

To begin, let us invert the Laplace transform presented in Eq. (\ref{CanZ}) to extract the density of states, i.e. 
\begin{equation}
\Omega(E) = \frac{1}{2 \pi i} \int_{a - i\infty}^{a + i\infty} Z(\beta) e^{\beta E} d\beta.
\end{equation} Now, defining \(S(\beta) = \ln Z(\beta) + \beta E\), we have 
\begin{equation}\label{OmegaCan}
\Omega(E) = \frac{1}{2 \pi i} \int_{a - i\infty}^{a + i\infty}e^{S(\beta)} d\beta,
\end{equation} where \(S(\beta)\) is the `entropy function'. Let us note that this entropy function describes the off-shell entropy at any temperature \(\beta\). If the equilibrium temperature is \(\beta_0\), then \(S(\beta_0) = S_0\) describes the equilibrium (on-shell) entropy\footnote{On-shell, the entropy function becomes \(S(\beta_0) = \ln Z(\beta_0) + \beta_0 E\). Using \(\beta_0 F = - \ln Z(\beta_0)\), we get \(F = E - TS\) (written without the subscripts). It should be remarked that while writing thermodynamic quantities and response functions, we will often omit the subscript `0' for the sake of brevity.}. However, the system can have small thermal fluctuations in its temperature about \(\beta_0\). To account for this, we expand \(S(\beta) = S_0 + (\partial_\beta^2 S)_0 (\beta - \beta_0)^2 + \cdots\), where \((\partial_\beta^2 S)_0\) is the second derivative of \(S(\beta)\) evaluated at \(\beta = \beta_0\). Substituting this expansion (truncated to second order because fluctuations are small) into Eq. (\ref{OmegaCan}) and performing the integral gives 
\begin{equation}
\Omega(E) = \frac{e^{S_0}}{\sqrt{2 \pi (\partial_\beta^2 S)_0}},
\end{equation} which means the microcanonical entropy \(S_{\rm mc} := \ln \Omega(E)\) is just
\begin{equation}\label{Smicrocan}
S_{\rm mc} = S_0 - \frac{1}{2} \ln \big[(\partial_\beta^2 S)_0\big] + \cdots,
\end{equation} where we ignore the constant terms. To summarize, the effect of fluctuations in the inverse temperature about its equilibrium value gives logarithmic corrections to the microcanonical entropy. These corrections are very general and hold for any system described by the canonical ensemble. Now one can show that \((\partial_\beta^2 S)_0 = (\Delta E)^2\), i.e. it is the variance of the internal energy and further, it is a simple exercise to show that \((\Delta E)^2 = T^2 C_{X^1, X^2, \cdots}\), where \(C_{X^1,X^2, \cdots}\) is the specific heat at constant values of \(X^1, X^2, \cdots\). Thus, just by knowing the form of \(C_{X^1,X^2, \cdots}\) at equilibrium, one can compute the logarithmic corrections appearing in the microcanonical entropy. It is remarkable that to achieve this, one doesn't need to know the exact form of the microscopic density of states, which is often not known for black holes. 

\subsection{Ensemble with two fluctuating coordinates}
Let us consider a situation where, in addition to a thermostat, we have a bath for some extensive variable \(X\). There are other extensive thermodynamic variables fixed by the boundaries of the system but we do not refer to them explicitly. In this situation, the partition function can be shown to take the following form:
\begin{equation}\label{2varZ}
\mathcal{Z} = \int \int \Omega(E,X) e^{-\beta E - xX} dE dX,
\end{equation} where \(x\) is some appropriate intensive parameter conjugate to \(X\). Examples include the grand canonical ensemble \cite{callen} where \(X\) can take the role of number of particles \(N\) for a gas, or electric charge \(Q\)/angular momentum \(J\) for black holes. For hydrostatic systems, and in the black hole chemistry framework \cite{open,CGQ,ACSHol}, yet another ensemble of the type described above is the isothermal-isobaric ensemble \cite{isob1,isob2} where \(X = V\) (thermodynamic volume) while \(x = \beta P\). From Eq. (\ref{2varZ}), we can extract the density of states as 
\begin{eqnarray}
\Omega(E,X) &=& \frac{1}{(2 \pi i)^2}  \int_{a - i\infty}^{a + i\infty} \int_{b - i\infty}^{b + i\infty} \mathcal{Z}(\beta,x) e^{\beta E +  xX} d\beta dx \nonumber \\
&=& \frac{1}{(2 \pi i)^2}  \int_{a - i\infty}^{a + i\infty} \int_{b - i\infty}^{b + i\infty} e^{S(\beta,x)} d\beta dx,
\end{eqnarray} where we have used the fact that \(S(\beta,x) = \ln \mathcal{Z}(\beta,x) + \beta E + xX\). One may consider a general situation, where there are fluctuations in the inverse temperature as well as in the value of \(x\). Thus, as in the previous subsection, we can expand the general entropy function \(S(\beta,x)\) about the equilibrium point (at which its first derivatives vanish) \((\beta_0,x_0)\). Then, after performing the integrals, we end up with the following expression for the corrected microcanonical entropy:
\begin{equation}\label{2coorS}
S_{\rm mc} = S_0 - \frac{1}{2} \ln D + \cdots,
\end{equation} where once again we have neglected constant terms and \(D = (\partial_\beta^2 S)_0  (\partial_x^2 S)_0 - [(\partial_\beta \partial_x S)_0]^2\) is the one-loop determinant in the statistical mechanics sense. It may be shown that \((\partial_\beta^2 S)_0  = (\Delta E)^2\) (as before), \((\partial_x^2 S)_0 = (\Delta X)^2 \), and \((\partial_\beta \partial_x S)_0  = \langle \Delta E \Delta X \rangle\), meaning that \(D\) is the determinant of the covariance matrix \cite{Chatterjee,open,CGQ}. Some straightforward computations give [see appendix-(\ref{appA})]
\begin{equation}\label{D2}
D = -T^2 \bigg( \frac{\partial X}{\partial x}\bigg)_T C_X,
\end{equation}
where \(C_X\) is the specific heat at fixed \(X\). Notice the negative sign appearing above. We will discuss two examples below.

\subsubsection{Grand canonical ensemble}\label{GCEsec}
In this case, \(X\) can be taken to correspond to some generic charge \(N\) such as electric charge or angular momentum. Then, \(x = - \beta \mu\) and from Eq. (\ref{D2}), one gets the following expression:
\begin{equation}
  D = T^3 \chi_T C_N,
\end{equation}
where \(C_N\) is the specific heat at constant \(N\), and \(\chi_T\) is the isothermal susceptibility associated with \(N\), defined as
\begin{equation}
\chi_T = \bigg( \frac{\partial N}{\partial \mu} \bigg)_T.
\end{equation}

\subsubsection{Isothermal-isobaric ensemble}
In this case, there is a bath for the volume, i.e. a barostat and the system reaches a state of thermodynamic equilibrium at fixed temperature and pressure \cite{isob1,isob2}. Then \(X = V\) and \(x = \beta P\) which means from Eq. (\ref{D2}), we have the following expression for \(D\) \cite{CGQ}:
\begin{equation}\label{DTP1}
D = T^3 V \kappa_T C_V,
\end{equation} where \(C_V\) is the specific heat at constant volume \(V\), and \(\kappa_T\) is the isothermal compressibility, defined as
\begin{equation}
\kappa_T = - \frac{1}{V} \bigg(\frac{\partial V}{\partial P} \bigg)_T.
\end{equation}

\subsection{Isoenthalpic-isobaric ensemble}
Let us now consider the isoenthalpic-isobaric ensemble which has not been discussed so far in the existing literature on logarithmic corrections to black hole entropy. The setting corresponds to the system reaching an equilibrium in contact with a barostat. Thus, the quantities \(P\) (fixed by barostat) and \(H = E +PV\) (fixed by walls of container) define the state of equilibrium. As we shall argue later, this ensemble is the analogue of the microcanonical ensemble in the framework of black hole chemistry [section-(\ref{ExtSec})]. \\

As far as we are aware, the isoenthalpic-isobaric ensemble \cite{nph11,nph12,nph22,nph33} has not been discussed extensively even in statistical mechanics literature. However, it finds application in molecular dynamics simulations \cite{nph1,nph2}. We will follow the framework put up in \cite{nph12,nph22} in our subsequent discussion. From a statistical mechanics viewpoint, if \(\mathcal{H}\) is the Hamiltonian describing the dynamics of the underlying system, then the density of microstates is defined as
\begin{equation}
\omega(H,P) = \int \delta[\mathcal{H} - (E + PV)] d\Gamma,
\end{equation} where \(d\Gamma\) is a suitable phase space measure. In the thermodynamic limit, the entropy is defined as \(S(H,P) = \ln \omega(H,P)\), which satisfies the first law:
\begin{equation}
dS = \beta dH - \beta V dP,
\end{equation} implying that temperature is defined as
\begin{equation}
\beta = \frac{1}{T} := \bigg(\frac{\partial S}{\partial H}\bigg)_{P}.
\end{equation} One can similarly find volume as
\begin{equation}
V := - \frac{1}{\beta}  \bigg(\frac{\partial S}{\partial P}\bigg)_{H}. 
\end{equation}
These relations are analogous, yet distinct from those of the microcanonical ensemble, where the entropy is defined as \(S(E,V) = \ln \Omega (E,V)\), and satisfies \(dS = \beta dE + \beta P dV\). A crucial difference between the microcanonical ensemble and the isoenthalpic-isobaric ensemble is that in the latter, neither energy \(E\) nor volume \(V\) is fixed (as long as the combination \(H = E + PV\) is fixed). Thus, one can have fluctuations in both internal energy and volume in this ensemble \cite{nph12}. \\

\subsubsection{Connection with microcanonical ensemble}
The connection between the isoenthalpic-isobaric entropy and the microcanonical entropy can be described as follows. Since \(S(E,V) = \ln \Omega(E,V)\) describes the microcanonical entropy and \(S(H,P) = \ln \omega (H,P)\) describes the isoenthalpic-isobaric entropy, then clearly they satisfy the first laws:
\begin{equation}\label{fl2s}
dS(E,V) = \beta dE + \beta P dV, \hspace{7mm} dS(H,P) = \beta dH - \beta V dP.
\end{equation} Although the first laws in the energy representation, i.e. \(dE = TdS - PdV\) and \(dH = TdS + VdP\) are connected by the Legendre transform \(H = E + PV\), the ones appearing in Eq. (\ref{fl2s}) are not related by a simple Legendre transform. This is because the degeneracies \(\Omega(E,V)\) and \(\omega(H,P)\) are not related by a Laplace transform. However, we may relate them via the following three-step process. \\

The isoenthalpic-isobaric ensemble is related to the isothermal-isobaric ensemble via Laplace transform, i.e. 
\begin{equation}\label{LTP}
\Delta(\beta, P) = \int_0^\infty \omega(H,P) e^{-\beta H} dH. 
\end{equation} Similarly, the isothermal-isobaric partition function is related to the canonical partition function via the Laplace transform (see \cite{CGQ} and references therein):
\begin{equation}
\Delta(\beta, P) = \int_0^\infty Z(T,V) e^{-\beta PV} dV,
\end{equation} where we have disregarded constant factors. But using Eq. (\ref{CanZ}), we have 
\begin{equation}\label{abcdef}
\Delta(\beta, P) = \int_0^\infty  \int_0^\infty  \Omega(E,V) e^{-\beta(E +  PV)} dE dV . 
\end{equation} Therefore, combining Eqs. (\ref{LTP}) and (\ref{abcdef}), we can write \cite{nph12}
\begin{equation}\label{micrelis}
\omega(H,P) = \frac{1}{2 \pi i} \int_{a - i\infty}^{a + i\infty} e^{\beta H} d\beta \bigg( \int_0^\infty  \int_0^\infty  \Omega(E,V) e^{-\beta(E +  PV)} dE dV\bigg).
\end{equation}
Eq. (\ref{micrelis}) expresses the relationship between the isoenthalpic-isobaric and microcanonical degeneracies.

\subsubsection{Logarithmic corrections to isoenthalpic-isobaric entropy}\label{iicor}
We know that the fixed \((T,P)-\)ensemble can be achieved from the isoenthalpic-isobaric ensemble by the use of a Laplace transform [Eq. (\ref{LTP})]. This means
\begin{equation}
\omega(H,P) = \frac{1}{2 \pi i} \int_{a - i\infty}^{a + i\infty} \Delta(\beta, \beta P) e^{\beta H} d\beta.
\end{equation} In the `one-loop' approximation, one gets the following expression for the isoenthalpic-isobaric density of states:
\begin{equation}
\omega(H,P) = \frac{e^{S_0}}{\sqrt{2 \pi (\partial^2_\beta S)_0}}.
\end{equation} Now, consider the equality
\begin{equation}\label{lkj}
\frac{\partial^2 S}{\partial \beta^2} \bigg|_{\beta = \beta_0} = \frac{\partial^2 \ln \Delta}{\partial \beta^2} \bigg|_{\beta = \beta_0},
\end{equation} where \(S(\beta) = \ln \Delta (\beta) + \beta H\) has been used. In order to find out the right-hand side, let us note that
\begin{equation}
\frac{\partial \ln \Delta}{\partial \beta}\bigg|_{\beta = \beta_0} = -H,
\end{equation} meaning that at equilibrium, the right hand side of Eq. (\ref{lkj}) is just \(-\partial H/\partial \beta\) at fixed \(P\). Thus, 
\begin{equation}\label{lkjh}
\frac{\partial^2 S}{\partial \beta^2} \bigg|_{\beta = \beta_0} = T^2 C_P,
\end{equation} where \(C_P = (\partial H/\partial T)_P\). Putting it all together gives the corrected isoenthalpic-isobaric entropy as 
\begin{equation}\label{isoiso}
S_{\rm isoenthalpic-isobaric} = S_0 - \frac{1}{2} \ln (T^2 C_P) + \cdots .
\end{equation}
This is analogous to Eq. (\ref{Smicrocan}) for the corrected microcanonical entropy with the only difference that presently \(C_P\) is the specific heat at fixed pressure (an intensive quantity), while in Eq. (\ref{Smicrocan}), the specific heat is evaluated at fixed extensive quantities like volume and number of particles.


\section{Logarithmic corrections to black hole entropy}\label{BHsec}
In this section, we will apply the general techniques discussed in the previous section to some black holes at non-zero (preferably high!) temperatures. In the subsequent subsections, we will set \(G_d = 1\) (Newton's constant) and the cosmological constant \(\Lambda\) shall be treated as a constant parameter. Then, in the next section, we will relax this and consider a framework where the cosmological constant leads to a bulk pressure.

\subsection{BTZ black holes}\label{BTZsubsec}
As our first example, let us consider the BTZ black hole which solves the Einstein equations in \((2+1)\) dimensions in the presence of a negative cosmological constant. The black hole mass \(M\), which is thermodynamically equivalent to the internal energy, can be expressed as \cite{BTZ}
\begin{equation}\label{Mbtz}
M: = E = \frac{S_{\rm BH}^2}{2 \pi^2 l^2} + \frac{\pi^2 J^2}{128 S_{\rm BH}^2},
\end{equation} where \(l\) denotes the AdS length scale, i.e. \(\Lambda = -1/l^2\) is the cosmological constant, \(J\) is the angular momentum, and \(S_{\rm BH} \sim r\), where \(r\) is the black hole size parameter (here, just the event horizon radius). The thermodynamic variables \((M,J)\) describe the system completely and one can write \(S_{\rm BH} = S_{\rm BH} (M,J)\) or alternatively, \(M = M (S_{\rm BH}, J)\) [Eq. (\ref{Mbtz})], to describe the fundamental thermodynamic relation\footnote{A fundamental thermodynamic relation describes a suitable potential function \(\Psi = \Psi(X^i)\) satisfying a first law, i.e. \(d\Psi = Y_i dX^i\). Consequently, \(Y_i = \frac{\partial \Psi}{\partial X^i}\) are the `equations of state' which describe the on-shell thermodynamic behavior of the system.}. Thus, the first law of black hole thermodynamics is \(\delta M := \delta E = T_{\rm H} \delta S_{\rm BH} + \Omega \delta J\), where \(T_{\rm H}\) is the Hawking temperature and \(\Omega\) is the angular velocity (notation not to be confused with number of microstates) conjugate to \(J\). The Hawking temperature is computed to be\footnote{In a microcanonical sense, where \(M =E\), the Hawking temperature is just \begin{equation}\label{Th}
T_{\rm H} = \bigg(\frac{\partial S_{\rm BH}}{\partial E}\bigg)^{-1}_{J},
\end{equation} where \(S_{\rm BH} = S_{\rm BH}(E,J)\) is the entropic fundamental thermodynamic relation.}
\begin{equation}
T_{\rm H} = \bigg( \frac{\partial E}{\partial S_{\rm BH}} \bigg)_{J} = \frac{S_{\rm BH}}{\pi^2 l^2} - \frac{\pi^2 J^2}{64 S_{\rm BH}^3} ,
\end{equation} and may go to zero if \(J \neq 0\) (the extremal point). In subsequent discussions, we will always consider \(T_{\rm H} > 0\) or equivalently, \(|J| < 8 S_{\rm BH}^2/ \pi^2 l\). The high-temperature limit is obtained as \(S_{\rm BH} \rightarrow \infty\) for which \(T_{\rm H} \sim S_{\rm BH} \sim r\). The specific heat at constant \(J\) is obtained straightforwardly and has the expression:
\begin{equation}
C_J = T_{\rm H} \bigg(\frac{\partial S_{\rm BH}}{\partial T_{\rm H}} \bigg)_{J} = \frac{64 S_{\rm BH}^5 - \pi^4 J^2 l^2 S_{\rm BH}}{64 S_{\rm BH}^4 + 3 \pi^4 J^2 l^2}.
\end{equation} It goes to zero at the extremal point. It is clear that in the high-temperature limit, the specific heat goes as \(C_J \sim S_{\rm BH} \sim r\). We discuss logarithmic corrections to the microcanonical entropy in two different ensembles below. 

\subsubsection{Energy fluctuations}
Following the prescription discussed in subsection-(\ref{cansec}), one can compute the logarithmic corrections to the microcanonical entropy due to fluctuations in energy. Since the variance of energy (mass) is \((\Delta E)^2 = T_{\rm H}^2 C_J\), while we have \(C_J > 0\) for \(T_{\rm H} > 0\) with no divergences (critical point), one can safely apply the statistical mechanics framework discussed in subsection-(\ref{cansec}) to obtain the log corrections. They take the form:
\begin{eqnarray}
S_{\rm mc} &=& S_{\rm BH} - \frac{1}{2} \ln (T_{\rm H}^2 C_{J}) + \cdots \nonumber \\
&\approx& S_{\rm BH} - \frac{1}{2} \ln S_{\rm BH} - \frac{1}{2} \ln \Bigg[ \frac{(64 S_{\rm BH}^4 - \pi^4 J^2 l^2 )^3}{4096 S_{\rm BH}^6 \pi^4 l^4 (64 S_{\rm BH}^4 + 3 \pi^4 J^2 l^2) } \Bigg]. 
\end{eqnarray}
These do in general differ from those obtained in \cite{Cardy} from Cardy's formula unless the one invokes the high-temperature limit, for which \cite{parthasarathi}
\begin{equation}\label{Sbtzcancor}
S_{\rm mc} = S_{\rm BH} - \frac{3}{2} \ln S_{\rm BH} + \cdots,
\end{equation}
matching with the coefficient found in \cite{Cardy}. 

\subsubsection{Energy + angular momentum fluctuations} 
Let us now discuss a scenario where the BTZ black hole equilibrates with respect to both a thermostat and a reservoir for angular momentum. Thus, the equilibrium state is controlled by the intensive parameters \(T_{\rm H}\) and \(\Omega\). In order to compute the logarithmic corrections to the microcanonical entropy, we must therefore consider the grand canonical ensemble discussed in subsection-(\ref{GCEsec}) to accommodate for both energy and angular momentum fluctuations, i.e. put  \(X = J\) and \(x = - \beta \Omega\). The simplified expression for \(D\) is [appendix-(\ref{appA})]
\begin{equation}
D =T_{\rm H}^3 \chi_T C_J,
\end{equation} where
\begin{equation}
\chi_T = \bigg( \frac{\partial J}{\partial \Omega} \bigg)_{T_{\rm H}},
\end{equation} is the isothermal `rotational' susceptibility. Although it is quite cumbersome to compute the susceptibility in the general case, one can consider the high-temperature limit where its dependence on \(S_{\rm BH}\) can be straightforwardly obtained. Let us begin by noting that the angular velocity is defined as
\begin{equation}
\Omega = \bigg( \frac{\partial E}{\partial J} \bigg)_{S_{\rm BH}} = \frac{\pi^2 J}{64 S_{\rm BH}^2}. 
\end{equation}
In the high-temperature limit, since \(T_{\rm H} \sim S_{\rm BH}\), it suffices to fix \(S_{\rm BH}\) in the above equation for fixing \(T_{\rm H}\), so that
\begin{equation}
\bigg( \frac{\partial \Omega}{\partial J} \bigg)_{T_{\rm H}} = \bigg( \frac{\partial \Omega}{\partial J} \bigg)_{S_{\rm BH}} \sim \frac{1}{S_{\rm BH}^2},
\end{equation} or equivalently, \(\chi_T \sim S_{\rm BH}^2\). Putting it all together gives the following logarithmic corrections to the microcanonical entropy:
\begin{eqnarray}
S_{\rm mc} &=& S_{\rm BH} - \frac{1}{2} \ln (T_{\rm H}^3 \chi_T C_J) + \cdots \nonumber \\
&\approx& S_{\rm BH} - 3 \ln S_{\rm BH}, \label{Sbtzmixcor}
\end{eqnarray} where we note that there is an extra contribution of \(-(3/2) \ln S_{\rm BH}\) as compared to Eq. (\ref{Sbtzcancor}), coming from the additional fluctuation coordinate \cite{Sarkar} (see also \cite{ACSHol}). This coefficient agrees with that obtained in \cite{Sen:2012dw} from microscopic analysis based on Cardy's formula. 

\subsection{Kerr-Newman black holes}
Next, let us consider the Kerr-Newman black hole which corresponds to a black hole of mass \(M\), carrying electric charge \(Q\), and spinning about an axis with angular momentum \(J\). Thus, the parameters \((M,Q,J)\) completely describe this particular solution of the Einstein-Maxwell's equations. The following fundamental relation describes the thermodynamics \cite{Davy}:
\begin{eqnarray}
M := E = \sqrt{2 S_{\rm BH} + (1/8 S_{\rm BH}) \bigg(J^2 + \frac{Q^4}{4}\bigg) + \frac{Q^2}{2}  },
\end{eqnarray} where \(S_{\rm BH}\) is the Bekenstein-Hawking entropy and goes with the black hole size parameter \(r\) as \(S_{\rm bh }\sim r^2\). Putting \(J = Q = 0\) leads to the Schwarzschild black hole, whereas putting \(J = 0\) gives rise to the Reissner-Nordstr\"om black hole. Since the first law of black hole thermodynamics in this case is \(\delta M := \delta E = T_{\rm H} \delta S_{\rm BH} + \Phi \delta Q  + \Omega \delta J \), one computes the Hawking temperature to be 
\begin{equation}
T_{\rm H} = \bigg( \frac{\partial E}{\partial S_{\rm BH}} \bigg)_{Q,J} = \frac{1}{E} \Bigg[1 - \frac{J^2 + \frac{Q^4}{4}}{16 S_{\rm BH}^2}\Bigg].
\end{equation}
The electric potential and angular velocity parameters are computed to be
\begin{eqnarray}
\Phi &=&  \bigg( \frac{\partial E}{\partial Q} \bigg)_{S_{\rm BH},J} = \frac{Q(Q^2 + 8 S_{\rm BH})}{16 E S_{\rm BH}}, \\
\Omega &=&  \bigg( \frac{\partial E}{\partial J} \bigg)_{S_{\rm BH},Q} =  \frac{J}{8 E S_{\rm BH}}.
\end{eqnarray}
One may now compute the logarithmic corrections to the microcanonical entropy from all the expressions above. Let us note that the specific heat at constant \(Q\) and \(J\) is given by
\begin{eqnarray}
C_{Q,J} &=& T_{\rm H} \bigg(\frac{\partial S_{\rm BH}}{\partial T_{\rm H}} \bigg)_{Q,J} = \frac{8 S_{\rm BH}^3}{J^2 + \frac{Q^4}{4} - 8 T_{\rm H}^2 S_{\rm BH}^3}\Bigg[1 - \frac{J^2 + \frac{Q^4}{4}}{16 S_{\rm BH}^2}\Bigg] \nonumber \\
&=&8 S_{\rm BH}^3 \times \frac{F_1 (S_{\rm BH}, Q, J)}{F_2 (S_{\rm BH}, Q, J)},
\end{eqnarray} where we have defined rational functions \(F_1\) and \(F_2\) as
\begin{equation}
F_1 (S_{\rm BH}, Q, J) = 1 - \frac{J^2 + \frac{Q^4}{4}}{16 S_{\rm BH}^2},
\end{equation}
\begin{equation}
F_2 (S_{\rm BH}, Q, J) = \bigg( J^2 + \frac{Q^4}{4}\bigg) - \frac{8 S_{\rm BH}^3  \Big[1 - \frac{J^2 + \frac{Q^4}{4}}{16 S_{\rm BH}^2}\Big]^2}{2 S_{\rm BH} + (1/8 S_{\rm BH}) \bigg(J^2 + \frac{Q^4}{4}\bigg) + \frac{Q^2}{2}} . 
\end{equation}

\subsubsection{Energy fluctuations}
Consider the case where the system is described by the canonical ensemble, i.e. the system reaches an equilibrium at temperature \(T_{\rm H}\) by exchanging matter (mass \(M\)) with a heat bath. Before computing the logarithmic corrections to entropy, let us comment on the behavior of the quantity \((\Delta E)^2\) (here, \(M\) is identified with energy \(E\)). It is easy to find the variance of the black hole mass since \(C_{Q,J}\) is known and it takes the following closed form expression: 
\begin{eqnarray}
(\Delta E)^2 = \frac{8 S_{\rm BH}^3}{\Big[2 S_{\rm BH} + (1/8 S_{\rm BH}) \big(J^2 + \frac{Q^4}{4}\big) + \frac{Q^2}{2}\Big]}  \frac{\Bigg[1 - \frac{J^2 + \frac{Q^4}{4}}{16 S_{\rm BH}^2}\Bigg]^3}{ \big( J^2 + \frac{Q^4}{4}\big) - \frac{8 S_{\rm BH}^3  \Big[1 - \frac{J^2 + \frac{Q^4}{4}}{16 S_{\rm BH}^2}\Big]^2}{2 S_{\rm BH} + (1/8 S_{\rm BH}) \big(J^2 + \frac{Q^4}{4}\big) + \frac{Q^2}{2}}}.
\end{eqnarray}
For \(Q = J = 0\), one finds that the variance is negative, indicating an inconsistency. So we shall never consider this limit in subsequent discussions. It is interesting to note that the variance of energy blows up at the Davies phase transition point \cite{Davy}. Physically, it seems plausible because fluctuations are very large near the critical point and therefore it doesn't make sense to compute logarithmic corrections at the critical point. Hence, we will restrict to the region where the specific heat \(C_{Q,J}\) is positive definite and finite, without worrying about the other side of the Davies line. \\

 Following the treatment presented in subsection-(\ref{cansec}), the logarithmic corrections are given by
\begin{eqnarray}
S_{\rm mc} &=& S_{\rm BH} - \frac{1}{2} \ln (T_{\rm H}^2 C_{Q, J}) + \cdots \nonumber \\
&\approx& S_{\rm BH} - \frac{3}{2} \ln S_{\rm BH} + \ln E - \frac{3}{2} \ln F_1 (S_{\rm BH}, Q, J) + \frac{1}{2} \ln F_2 (S_{\rm BH}, Q, J). 
\end{eqnarray}
Although the corrections have a complicated functional form, let us note that there is a term of the type [Eq. (\ref{loggen})] \(\ln S_{\rm BH}\) with \(k = 3/2\), the same coefficient as obtained for BTZ black holes [Eq. (\ref{Sbtzcancor})]. We remind the reader than unlike the case of BTZ black holes, or higher-dimensional black holes in AdS [subsection-(\ref{AdSNonextend})], for Kerr-Newman black holes asymptotic to flat space, it doesn't make sense to consider a `high-temperature' limit in which the corrections although more pronounced, are expected to have a simplified form. This is because in the absence of a negative cosmological constant, there is no large black hole solution which gives large Hawking temperature for large entropy. Instead, for Kerr-Newman black holes, a high-temperature limit implies a negative specific heat at constant \(Q\) and \(J\) leading to instability.

\subsection{Black holes in AdS spacetimes}\label{AdSNonextend}
We now consider black holes in AdS spacetimes, i.e. in the presence of a negative cosmological constant \(\Lambda = - \frac{(d-1)(d-2)}{2l^2}\), where \(d\) is the number of spacetime dimensions and \(l\) is the AdS length scale. For definiteness, we consider \(d=4\) and electrically charged black holes in AdS for which the mass is given by \cite{Chamblin,Chamblin1}
\begin{equation}\label{AdSNonextendmass}
M = \frac{r_+}{2} \bigg( 1 + \frac{r_+^2}{l^2} + \frac{Q^2}{r_+^2} \bigg),
\end{equation} where \(r_+\) is the event horizon radius, i.e. the Bekenstein-Hawking entropy is \(S_{\rm BH} = \pi r_+^2\). Thus, in terms of \(S_{\rm BH}\) and \(Q\), which take the role of thermodynamic variables, the mass is expressed as
\begin{equation}
M := E = \sqrt{\frac{S_{\rm BH}}{4\pi}} \bigg( 1 + \frac{S_{\rm BH}}{\pi l^2} + \frac{ \pi Q^2}{S_{\rm BH}} \bigg),
\end{equation} describing the fundamental thermodynamic relation. This means we have the first law \(\delta M := \delta E = T_{\rm H} \delta S_{\rm BH} + \Phi \delta Q\), implying that the Hawking temperature and the electric potential are 
\begin{eqnarray}
T_{\rm H} &=& \bigg( \frac{\partial E}{\partial S_{\rm BH}} \bigg)_{Q} = \frac{1}{\sqrt{16 \pi}} \Bigg[S_{\rm BH}^{-1/2} + \frac{3 S_{\rm BH}^{1/2}}{\pi l^2} - \frac{\pi Q^2}{S_{\rm BH}^{3/2}} \Bigg], \label{chargednonextendedhawking} \\
\Phi &=& \bigg( \frac{\partial E}{\partial Q} \bigg)_{S_{\rm BH}} = \frac{\sqrt{\pi} Q}{\sqrt{S_{\rm BH}}}. \label{Phinonextendedchargedads}
\end{eqnarray} It should be remarked here that one can consider an alternate scheme, where \(Q^2\) rather than \(Q\) is treated as a thermodynamic variable. This has been briefly discussed in appendix-(\ref{Q2}). Now, the specific heat at fixed electric charge is 
\begin{equation}\label{CQRNnonextended}
C_Q = T_{\rm H} \bigg(\frac{\partial S_{\rm BH}}{\partial T_{\rm H}} \bigg)_{Q} = 2 S_{\rm BH} \times \frac{S_{\rm BH} \pi l^2 + 3 S_{\rm BH}^2 - \pi^2 l^2 Q^2}{-S_{\rm BH} \pi l^2 + 3 S_{\rm BH}^2 + 3 l^2 \pi^2 Q^2},
\end{equation} which blows up at the critical point \cite{Chamblin}. With this setup, we first consider energy fluctuations. 

\subsubsection{Energy fluctuations} 
The variance of energy (mass) is \((\Delta E)^2 = T_{\rm H}^2 C_{Q}\) and blows up at the critical point. This is because the specific heat scales as \(C_Q \sim |T_{\rm H} - T_c|^{-1}\) near the critical temperature leading to an infinite discontinuity at \(T_{\rm H} = T_c\). So our corrections are well defined away from the critical point where fluctuations are small. The logarithmic corrections to the microcanonical entropy take the following form:
\begin{eqnarray}
S_{\rm mc} &=& S_{\rm BH} - \frac{1}{2} \ln (T_{\rm H}^2 C_{Q}) + \cdots \nonumber \\
&\approx& S_{\rm BH} + \ln S_{\rm BH} - \frac{3}{2} \ln (S_{\rm BH} \pi l^2 + 3 S_{\rm BH}^2 - \pi^2 l^2 Q^2) \nonumber \\ 
&+& \frac{1}{2} \ln (-S_{\rm BH} \pi l^2 + 3 S_{\rm BH}^2 + 3 l^2 \pi^2 Q^2) ,
\end{eqnarray}
where we have neglected some constants. Notice that the first two terms of these corrections are of the form suggested in Eq. (\ref{loggen}) with \(k = -1\). It is imperative to take the high-temperature limit, for which \(S_{\rm BH}\) is large and therefore, from Eqs. (\ref{chargednonextendedhawking}) and (\ref{CQRNnonextended}), one has \(T_{\rm H} \sim S_{\rm BH}^{1/2}\) and \(C_Q \sim S_{\rm BH}\). Thus, the logarithmic corrections are
\begin{equation}\label{abcd}
S_{\rm mc} = S_{\rm BH} - \ln S_{\rm BH} + \cdots,
\end{equation} which corresponds to Eq. (\ref{loggen}) with \(k = +1\). In \(d\) spacetime dimensions, one has \(T_{\rm H} \sim S_{\rm BH}^{1/(d-2)}\) and \(C_Q \sim S_{\rm BH}\), which means the corrections go as
\begin{equation}\label{abcde}
S_{\rm mc} = S_{\rm BH} - \frac{d}{2(d-2)} \ln S_{\rm BH} + \cdots,
\end{equation} which were also found in \cite{parthasarathi} for the high-temperature limit of Schwarzschild-AdS black holes. 

\subsubsection{Energy + charge fluctuations}
Consider the same system as before, i.e. the charged black hole in four-dimensional AdS but in the grand canonical ensemble, i.e. in the presence of a bath for the electric charge, which means the electric potential \(\Phi\) is fixed. Then, the logarithmic corrections are of the form given in Eq. (\ref{2coorS}) with \(D = T_{\rm H}^3 \chi_T C_Q\), where 
\begin{equation}\label{elecsus}
\chi_T = \bigg(\frac{\partial Q}{\partial \Phi}\bigg)_{T_{\rm H}},
\end{equation} is the isothermal charge susceptibility. \\

We consider the high-temperature limit where \(S_{\rm BH} \sim T_{\rm H}^2\), implying that for imposing fixed \(T_{\rm H}\), it suffices to fix \(S_{\rm BH}\) in Eq. (\ref{elecsus}). Thus, we can write
\begin{equation}
\bigg( \frac{\partial \Phi}{\partial Q} \bigg)_{T_{\rm H}} = \bigg( \frac{\partial \Phi}{\partial Q} \bigg)_{S_{\rm BH}} \sim \frac{1}{\sqrt{S_{\rm BH}}},
\end{equation} where we have used Eq. (\ref{Phinonextendedchargedads}). This gives \(\chi_T \sim \sqrt{S_{\rm BH}}\) and the corrections read
\begin{equation}\label{GCEcharged1}
S_{\rm mc} = S_{\rm BH} - \frac{3}{2} \ln S_{\rm BH} + \cdots,
\end{equation} where we note the appearance of the coefficient \(k = 3/2\) once again.\\

 A straightforward generalization to \(d\)-dimensional spacetimes is possible if we note that in the high-temperature limit, \(M := E \sim r^{d-1}\), where \(r\) is the size parameter (here, just the horizon radius \(r_+\)). Subsequently, because \(S_{\rm BH} \sim r^{d-2}\), we have the scaling \(E \sim S_{\rm BH}^{\frac{d-1}{d-2}}\), which means \(T_{\rm H} \sim S_{\rm BH}^{1/(d-2)}\) and \(C_Q \sim S_{\rm BH}\). Further, \(\Phi \sim Q r^{-(d-3)}\) gives \(\chi_T \sim S_{\rm BH}^{\frac{d-3}{d-2}}\), which means one gets logarithmic corrections of the following form:
 \begin{equation}
S_{\rm mc} = S_{\rm BH} - \frac{d-1}{d-2} \ln S_{\rm BH} + \cdots.
\end{equation}
It is clear that different ensembles lead to different logarithmic corrections stemming from different choices of fluctuation coordinates. These corrections are of thermal origin and shall arise when non-extremal black holes are considered in different Gibbs ensembles. In this respect, the corrections described in this section are quite robust.

\section{Logarithmic corrections in black hole chemistry}\label{ExtSec}
In this section, we consider the framework of black hole chemistry where the cosmological constant leads to the notion of thermodynamic pressure \cite{Kastor:2009wy,Cvetic:2010jb,Dolan:2010ha,Dolan:2011xt,Dolan119,Kubiznak:2012wp,45}. The formal definition of pressure is
\begin{equation}\label{Pdef}
P = - \frac{\Lambda}{8 \pi G_d},
\end{equation} in \(d\) dimensions. One can fix the Newton's constant to some value, say put \(G_d = 1\) and then, the variations of the cosmological constant manifest themselves as variations of pressure, i.e. \(\delta P \sim \delta \Lambda\) in the first law of black hole thermodynamics. Thus, the first law is augmented with a novel pressure-volume term \cite{Kastor:2009wy}, i.e. \(\delta M = T_{\rm H} \delta S_{\rm BH} + V \delta P + \Phi \delta Q + \Omega \delta J\), and this leads to the definition of a conjugate volume \cite{Kastor:2009wy,Cvetic:2010jb}:
\begin{equation}\label{Vdef}
V = \bigg(\frac{\partial M}{\partial P}\bigg)_{S_{\rm BH}, Q, J}.
\end{equation}
A consequence of this new first law with a `\(V \delta P\)' term suggests that one should interpret the black hole mass to be the enthalpy \(H\) of the spacetime \cite{Kastor:2009wy}, rather than its internal energy \(E\). Simple scaling arguments reveal that the Smarr formula reads \(M = 2TS - 2PV + \Phi Q\) for the four-dimensional charged black hole in AdS [section-(\ref{AdSNonextend})] and can be appropriately generalized to higher dimensions even in the presence of angular momentum \cite{45}. This framework, where black hole thermodynamics is supplemented with `chemical variables' \(P\) and \(V\) has been termed as black hole chemistry.  \\

If we combine Eqs. (\ref{Pdef}) and (\ref{Vdef}) with Eq. (\ref{AdSNonextendmass}), it gives \(V = \frac{4 \pi r_+^3}{3}\), which coincides with the naive geometric volume, i.e. the volume of a sphere of radius \(r_+\) in AdS \cite{Kastor:2009wy,Cvetic:2010jb,Dolan:2010ha,Dolan:2011xt,Dolan119}. This can be shown to be true for a wide variety of black holes such as the BTZ black hole in the absence of electric charge and various non-rotating black holes in higher dimensions. An immediate consequence of this is that entropy and volume are not independent, giving \(C_V =0\) for all such systems. The situation is similar to that discussed in appendix-(\ref{Q2}) in the sense that \(V\) is independent of \(P\), meaning that the Legendre transform \(E = H- PV\) (where \(M := H\)) is not well defined. In other words, the microcanonical ensemble and associated thermodynamic functions are not well defined and require some regularization. However, this excludes various other black holes, most notably Kerr-AdS\footnote{Other well known examples are charged BTZ, Kerr-Newman-AdS, and STU black holes \cite{Johnsonc}. However, charged BTZ black holes violate the reverse isoperimetric inequality and have negative specific heat at constant volume \cite{Johnsonbtz}.}, for which \(V\) is no longer the naive geometric volume but exceeds it, nevertheless satisfying the reverse isoperimetric inequality \cite{Cvetic:2010jb}.\\

\subsection{Logarithmic corrections to isoenthalpic-isobaric entropy}
In section-(\ref{BHsec}), we have discussed the microcanonical entropy for black holes and the logarithmic corrections it receives due to thermodynamic fluctuations. However, as we shall now argue, it is the isoenthalpic-isobaric entropy, rather than the microcanonical entropy, that fits in more naturally within the framework of black hole chemistry. Let us recall the first law of thermodynamics in black hole chemistry, i.e. \(\delta H = T_{\rm H} \delta S_{\rm BH} + V \delta P + \cdots\), where we have \(M := H\), and the fundamental thermodynamic relation is of the generic form \(H = H(S_{\rm BH}, P, Q, J)\). Since temperature is positive, this can be inverted to give \(S_{\rm BH} = S_{\rm BH}(H, P, Q, J)\). Now, because just the parameters \((M,\Lambda, Q, J)\) specify the black hole solution completely, the entropic fundamental relation \(S_{\rm BH} = S_{\rm BH}(H, P, Q, J)\) contains all the information about the thermodynamics. This corresponds to the entropic fundamental relation of the isoenthalpic-isobaric ensemble which is not defined at fixed \(E\) and \(V\) as with the microcanonical entropy\footnote{The isoenthalpic-isobaric ensemble naturally describes black holes in the framework of black hole chemistry. It is because the cosmological constant which appears directly in the black hole solution is just the pressure, i.e. any thermodynamic equilibrium state is defined by a given value of pressure (along with specifying other charges like mass, electric charge, angular momentum), while volume is a derived quantity. Moreover, the black hole mass is just the enthalpy, suggesting the applicability of the isoenthalpic-isobaric ensemble. In other words, the presence of cosmological constant is essentially analogous to having the system (here, black hole) in contact with a barostat.}. In other words, \(P\) is fixed at equilibrium (it directly comes from background cosmological constant) but \(E\) and \(V\) are free to vary with the constraint that \(H = E + PV\) stays fixed at a given equilibrium state. Considering all this, it seems pretty natural to associate an isoenthalpic-isobaric interpretation to the black hole entropy, with the entropic first law:
\begin{equation}
\delta S_{\rm BH} = \frac{\delta H}{T_{\rm H}} - \frac{V}{T_{\rm H}} \delta P - \frac{\Phi}{T_{\rm H}} \delta Q  - \frac{\Omega}{T_{\rm H}} \delta J ,
\end{equation} meaning that the Hawking temperature is defined as
\begin{equation}
T_{\rm H} = \bigg(\frac{\partial S_{\rm BH}}{\partial H}\bigg)^{-1}_{P,Q,J},
\end{equation} rather than the reciprocal of a derivative with respect to \(E\) as in Eq. (\ref{Th}). Therefore, instead of considering corrections to the microcanonical entropy, one should look out for computing corrections to the isoenthalpic-isobaric entropy \(S_{\rm isoenthalpic-isobaric}\), which is different from \(S_{\rm mc}\). Following the framework presented in subsection-(\ref{iicor}), one can compute such corrections straightforwardly. Below, we consider two examples. 

\subsubsection{BTZ black holes}
Let us consider BTZ black holes in the black hole chemistry framework. The volume scales with the entropy as \(V \sim S_{\rm BH}^2\). Further, at high temperatures, \(T_{\rm H} \sim S_{\rm BH}\), while \(\kappa_T\) does not scale with \(S_{\rm BH}\). Thus, Eq. (\ref{isoiso}) gives
\begin{equation}\label{extendedbtzcorr}
S_{\rm isoenthalpic-isobaric} \approx S_{\rm BH} - \frac{3}{2} \ln S_{\rm BH} ,
\end{equation} with a coefficient \(k = 3/2\) that matches exactly with the one obtained earlier \cite{Carlip,parthasarathi}. 

\subsubsection{Non-rotating black holes in AdS}
In the four-dimensional case and at high temperatures, we have \(T_{\rm H} \sim S^{1/2}_{\rm bh}\) and \(C_P \sim S_{\rm BH}\), which means we get from Eq. (\ref{isoiso}), the following corrections:
\begin{equation}
S_{\rm isoenthalpic-isobaric} \approx S_{\rm BH} - \ln S_{\rm BH} ,
\end{equation} with a coefficient \(k = 1\) that matches exactly with the result found earlier for four-dimensional black holes in AdS at high temperatures in \cite{parthasarathi}. In fact, one can check that the results match in any number of spacetime dimensions. To summarize, the isoenthalpic-isobaric entropy in black hole chemistry is the analogue of microcanonical entropy from formalisms where \(\Lambda\) variations are not considered. Although the corrections to the isoenthalpic-isobaric entropy obtained here match with those appearing in Eqs. (\ref{Sbtzcancor}) and (\ref{abcd}), it should be borne in mind that upon introducing \(P\) as a thermodynamic variable, one is considering a new pair of variables and therefore taking into account the possibility of new fluctuations. The essential reason behind the fact that the form of the corrections remain unchanged is that the scaling behavior of various thermodynamic quantities with the size parameter \(r\) of the black hole remains unchanged. Below, we discuss logarithmic corrections to the microcanonical entropy\footnote{Such corrections were studied earlier in the isothermal-isobaric and open ensembles \cite{open,CGQ}.}.  As we shall see, these corrections will be different from the ones obtained above or in section-(\ref{BHsec}).

\subsection{Logarithmic corrections to microcanonical entropy}\label{extmicregsec}
If one considers a situation where there are energy fluctuations only, one can use Eq. (\ref{Smicrocan}) to compute the corrections to the microcanonical entropy due to energy fluctuations. For Kerr-AdS black holes, the result is of the form \cite{CGQ}: \(S_{\rm mc} \approx S_{\rm BH} + \ln S_{\rm BH} - \frac{1}{2} \ln F_3(S_{\rm BH}, P, J) \) where \(F_3(S_{\rm BH}, P, J)\) is a rational function of its arguments. The first two terms resemble Eq. (\ref{loggen}) for \(k = -1\). Similarly, if one would consider a situation where there are both energy and volume fluctuations, then the corrections to the microcanonical entropy are obtained by computing the expression for \(D\) given in Eq. (\ref{DTP1}). The final result is of the form \(S_{\rm mc} \approx S_{\rm BH} + \ln S_{\rm BH} - \frac{1}{2} \ln F_4(S_{\rm BH}, P, J) \) where \(F_4(S_{\rm BH}, P, J)\) is another rational function (different from \(F_3\)) of its arguments. Notice that remarkably, the coefficient \(k = -1\) has remained unchanged \cite{CGQ}. \\

However, these corrections are not well defined if one puts \(J \rightarrow 0\) for which one has a Schwarzschild-AdS black hole with \(C_V = 0\). These black holes saturate the reverse isoperimetric inequality and their thermodynamic volume coincides with the geometric volume \cite{Cvetic:2010jb}. As mentioned earlier, the microcanonical entropy is not a well-defined function in this setting, given that one starts in a fixed pressure ensemble and the Legendre transform used to map to the fixed volume ensemble is singular. This points towards the existence of constraints\footnote{The relation between \(S_{\rm BH}\) and \(V\), independent of \(P\), can be treated as a constraint.}, which have a microscopic origin, i.e. it is not possible to thermally excite the degrees of freedom at a fixed thermodynamic volume. Now, we define a regularized microcanonical density of states:
\begin{equation}
\Omega^\lambda (E,V,Q) := \lambda \Omega (E,V,Q),
\end{equation} where \(\lambda \in [0,1]\) is a scale factor intrinsic to the system. Then, the microcanonical entropy takes the form \(S^\lambda_{\rm mc} = S_{\rm mc} + \ln \lambda\) meaning that the corrections given in Eq. (\ref{Smicrocan}) from the canonical ensemble become
\begin{equation}\label{canreg}
S^\lambda_{\rm mc} = S_{\rm BH} - \frac{1}{2} \ln T_{\rm H}^2  + \ln \frac{\lambda^2}{C_V}.
\end{equation}
To this end, we can consider Schwarzschild-AdS as well as charged black holes in AdS as the limit \(C_V \rightarrow 0^+\) together with the condition \(\lambda \rightarrow 0^+\) such that \(\lim_{\lambda,C_V \rightarrow 0^+} (\lambda^2/C_V)\) is finite. Although this regularization scheme looks rather artificial, we should note that since for these black holes, \(C_V = 0\) (constant) intrinsically, it does not have any dependence on the black hole size and therefore its removal cannot affect the final result\footnote{In other words, since \(C_V\) is a constant, we can choose the constant \(\lambda \sim \sqrt{C_V}\) and this regularizes the corrections.}, i.e. the coefficient \(k\) appearing in Eq. (\ref{loggen}). In this sense, the regularization performed above gives the log corrected microcanonical entropy for black holes which saturate the reverse isoperimetric inequality, independent of \(\lambda\) which is just a constant and the ratio \(\lambda/C_V^2\) can be ignored along with other constants. In the isothermal-isobaric ensemble, going along the same lines as before, the corrected and regularized microcanonical entropy reads
\begin{equation}\label{isothermreg}
S^\lambda_{\rm mc} = S_{\rm BH} - \frac{1}{2} \ln (V T_{\rm H}^3 \kappa_T)  + \ln \frac{\lambda^2}{C_V},
\end{equation} where symbols have their usual meanings. In further discussions, we will abuse notation by dropping the superscript \(\lambda\) and it will be understood that one is talking about the regularized microcanonical entropy. We now consider two examples below. 

\subsubsection{BTZ black holes}
Since we have \(T_{\rm H} \sim S_{\rm BH}\) and \(C_P \sim S_{\rm BH}\), we get the following corrections in the isothermal-isobaric ensemble [Eq. (\ref{isothermreg})] \cite{CGQ}:
\begin{equation}
S_{\rm mc} \approx S_{\rm BH} - \frac{5}{2} \ln S_{\rm BH}  {\rm ~~} \leftarrow {\rm ~~(Energy~+~volume~fluctuations)},
\end{equation} which are distinct from those obtained earlier in subsection-(\ref{BTZsubsec}). Now, if one were to consider energy fluctuations alone, i.e. the canonical ensemble, then the corrections read [Eq. (\ref{canreg})]
\begin{equation}\label{lll}
S_{\rm mc} \approx S_{\rm BH} - \ln S_{\rm BH} {\rm ~~} \leftarrow {\rm ~~(Energy~fluctuations)},
\end{equation} which give \(k = 1\) as opposed to \(k = 3/2\) found in Eq. (\ref{Sbtzcancor}). It might have happened due to the fact that in Eq. (\ref{Sbtzcancor}), the logarithmic corrections originated from the fluctuations of mass \((M := E)\) whereas the corrections appearing in Eq. (\ref{lll}) are fluctuations in \(E = M - PV\), where clearly the contribution from some degrees of freedom (\(\sim PV\)) has been subtracted out. It can be argued that for \(C_V = 0\), the fluctuations in \(E\) and \(V\) are not independent because the covariance matrix is singular, i.e. the covariance of \(E\) and \(V\) is equal to product of the standard deviations of the variables. Mathematically, the mismatch happens due to the fact that \(C_V\) is a constant and does not scale with \(S_{\rm BH}\). 

\subsubsection{Non-rotating black holes in AdS}
For black holes asymptotic to AdS, at high temperatures, i.e. for large sizes of the large black hole branch \((r_+ >> l)\), the entropy and temperature scale as \(S_{\rm BH} \sim T_{\rm H}^2\) in four dimensions, while \(V \sim S_{\rm BH}^{3/2}\). The isothermal compressibility does not scale with \(S_{\rm BH}\) at high temperatures and therefore, Eq. (\ref{isothermreg}) gives the following logarithmic corrections to the microcanonical entropy, starting from the isothermal-isobaric ensemble:
\begin{equation}
S_{\rm mc} \approx S_{\rm BH} - \frac{3}{2} \ln S_{\rm BH} {\rm ~~} \leftarrow {\rm ~~(Energy~+~volume~fluctuations)}.
\end{equation}
On the other hand, considering energy fluctuations alone, we have the following corrected microcanonical entropy from Eq. (\ref{canreg}):
\begin{equation}\label{ggg}
S_{\rm mc} \approx S_{\rm BH} - \frac{1}{2} \ln S_{\rm BH} {\rm ~~} \leftarrow {\rm ~~(Energy~fluctuations)},
\end{equation} 
where the coefficient in front of the logarithm differs from that of Eq. (\ref{abcd}). One can go further and consider logarithmic corrections to the microcanonical entropy of charged black holes in AdS due to energy and charge fluctuations, keeping volume fixed. The result is \(k = 1\) which is once again different from that obtained in Eq. (\ref{GCEcharged1}). We remind the reader that this mismatch and that between Eqs. (\ref{ggg}) and (\ref{abcd}) are due to the fact that \(C_V\) is independent of the black hole size. Generalization to \(d\) dimensions is straightforward and we do not pursue it here. 

\section{Holographic black hole chemistry}\label{holsec}
Let us consider a holographic setup for black hole chemistry. In the two subsections below, we explore two different proposals to the holographic setup. 

\subsection{Variable Newton's constant}\label{holsec1}
In a second possibility stemming from Eq. (\ref{Pdef}), one can also vary the Newton's constant \cite{visser,mann,zhao} (see also \cite{bose,K,Pcft}), meaning that the variations in bulk pressure go as \(\delta P \sim \delta (\Lambda/G_d)\). Since \(\Lambda\) and \(G_d\) can vary independently, one obtains two independent work terms in the first law. As has been argued in \cite{visser,mann}, this setup leads to a consistent matching between the first laws in the bulk and the boundary. Furthermore, variations of \(G_d\) ensure that the thermodynamics is truly extensive \cite{visser,zhao}, i.e. the fundamental thermodynamic relations such as \(M =M (S_{\rm BH}, \cdots)\) or \(S_{\rm BH} = S_{\rm BH} (M, \cdots)\) are homogenous functions of degree one in each extensive argument.\\

Prior to computing any logarithmic corrections, let us carefully choose the correct ensemble for the bulk, as well as, for the boundary. In the bulk, following \cite{mann}, the first law of thermodynamics for charged black holes in AdS reads
\begin{equation}\label{firstlawbulkhol}
\delta H = T_{\rm H} \delta S_{\rm BH} + V_c \delta P + \mu \delta c + \Phi \delta Q,
\end{equation} where \(H\) is the enthalpy (mass), \(P = - \frac{\Lambda}{ 8 \pi G_d}\) is the new pressure arising due to both \(\Lambda\) and \(G_d\) variations, while \(c \sim \frac{l^{d-2}}{G_d}\) is the central charge of the field theory to which the bulk gravity theory is dual to (\(c\) is large). Clearly, the first law above differs from that considered earlier. Let us first point out some new features below. \\

Since \(\Lambda\) and \(G_d\) can vary independently, one notices the existence of two independent work terms  \(V_c \delta P\) and \(\mu \delta c\) in Eq. (\ref{firstlawbulkhol}), as opposed to just a \(V \delta P\) term considered earlier. It should also be remarked that \(V_c\) is different from \(V\) considered earlier, because we are now taking into account variations in \(G_d\) too! A significant consequence of varying \(G_d\) is that now \(S_{\rm BH}\) and \(V_c\) turn out to be independent (because \(S_{\rm BH} \sim r^{d-2}/G_d\) is no longer just a function of \(r\)) and this means \(C_V \neq 0\), unlike in the setting considered earlier where \(C_V = 0\). Due to this, one can study the logarithmic corrections to the microcanonical entropy of black holes in this setting, without the need of regularizing the microcanonical density of states\footnote{Such corrections to the microcanonical entropy have been explored in \cite{ACSHol}.} as performed in subsection-(\ref{extmicregsec}).\\

Let us now consider the boundary. The boundary theory is a field theory, where the energy\footnote{We use the notation \(U\) for internal energy here, to distinguish from \(E\) used for energy on the gravity side which is \(E = M - PV_c\) for black hole chemistry and \(E = M\) for frameworks where \(\Lambda\) is fixed [section-(\ref{BHsec})].} \(U\) scales with the number of degrees of freedom \(c\), i.e. \(U \sim c\). The first law of thermodynamics is \cite{mann,visser}
\begin{equation}\label{firstlawboundhol}
\delta U = T \delta S - p \delta \mathcal{V} + \mu \delta c + \tilde{\Phi} \delta \tilde{Q},
\end{equation} where \(p\) and \(\mathcal{V}\) are the pressure and volume variables of the field theory respectively, distinct from \(P\) and \(V_c\) of the bulk \cite{Pcft}. The parameters \(\tilde{Q}\) and \(\tilde{\Phi}\) are related to their bulk counterparts as \cite{mann}
\begin{equation}
\tilde{Q} = \frac{Ql}{\sqrt{G_d}}, \hspace{10mm} \tilde{\Phi} = \frac{\sqrt{G_d} \Phi}{l}.
\end{equation}
If \(Z\) be the partition function of the field theory, \(\ln Z \sim c\), \(U = - \frac{\partial}{\partial \beta} \ln Z\), and \(p = \frac{1}{\beta} \frac{\partial}{\partial \mathcal{V}} \ln Z\), where \(\beta = 1/T\). Thus, the pressure of the field theory is determined from the field theory partition function by differentiation with respect to the volume which scales as\footnote{We consider the field theory to be on a compact space with the generic metric \(dl^2 = -dt^2 + l^2 d\Omega_{k,d-2}^2\), where \(k = +1, 0, -1\); respectively indicating towards spherical, flat, and hyperbolic topologies.} \(\mathcal{V} \sim l^{d-2}\), unlike the bulk pressure \(P \sim 1/G_d l^2\) which comes directly from the gravity solution. The black hole mass \(M\) in the bulk coincides with the internal energy \(U\) on the boundary, while the temperature and entropy of the field theory are related to those of the bulk \cite{mann,visser,bose,K,Pcft}. \\

Therefore, in going from bulk to boundary, the bulk first law [Eq. (\ref{firstlawbulkhol})] maps naturally to the boundary first law [Eq. (\ref{firstlawboundhol})]. Let us note that the bulk first law indicates that the system equilibrates at a constant pressure, while the boundary first law indicates that the system equilibrates at a fixed volume. Since this appears to be a consistent framework, one is led to associate the bulk first law with that of the isoenthalpic-isobaric ensemble, while the boundary first law corresponds to that of the microcanonical ensemble. This can be summarized by the following diagram below:

\begin{center}
 \begin{tikzcd}[row sep=huge,column sep=huge]
\delta E = T_{\rm H} \delta S_{\rm BH} + \Phi \delta Q \arrow[r, "S-{\rm function}"] \arrow[d, "{\rm Variations~of}~\Lambda"] 
& S_{\rm BH}(E,Q) \arrow[d, "{\rm Variations~of}~\Lambda"] \\
 \delta H = T_{\rm H} \delta S_{\rm BH} + V \delta P + \Phi \delta Q \arrow[r, "S-{\rm function}"] \arrow[d, "{\rm Variations~of}~G_d"] \arrow[u, bend left,  "{\rm Fixed}~\Lambda"]
& S_{\rm BH}(H,P,Q) \arrow[d, "{\rm Variations~of}~G_d"]  \arrow[u, bend left,  "{\rm Fixed}~\Lambda"]\\
\delta H = T_{\rm H} \delta S_{\rm BH} + V_c \delta P + \mu \delta c + \Phi \delta Q \arrow[d, "{\rm AdS/CFT~dictionary}"]  \arrow[r,  "S-{\rm function}"] \arrow[u, bend left,  "{\rm Fixed}~G_d"]
& S_{\rm BH}(H,P,c,Q)  \arrow[u, bend left,  "{\rm Fixed}~G_d"] \\
\delta U = T \delta S - p \delta \mathcal{V} + \mu \delta c + \tilde{\Phi} \delta \tilde{Q}  \arrow[r,  "S-{\rm function}"]
& S(U,\mathcal{V},c,\tilde{Q})
\end{tikzcd}
\end{center}
which also includes the larger framework involving all other cases considered so far (angular momentum not made explicit). Thus, one expects that the corrections to the isoenthalpic-isobaric entropy in the bulk should match with the corrections to the microcanonical entropy on the boundary. Let us discuss two cases below. 

\subsubsection{BTZ black holes}\label{BTZsubsub}
Let us consider fluctuations in \(H\) in the bulk for BTZ black holes. The logarithmic corrections to the isoenthalpic-isobaric entropy are given by Eq. (\ref{isoiso}). Thus, we are required to compute the Hawking temperature and the specific heat at constant pressure (+ constant angular momentum and central charge) to obtain the corrections. A straightforward computation reveals that 
\begin{equation}
T_{\rm H} :=\biggl(\frac{\partial H}{\partial S_{\rm BH}}\biggr)_{P,c,J} = \frac{\left(9 S_{\rm BH}^4-4 \pi ^4 c^2 J^2\right) \sqrt{\frac{4 \pi ^4 c^2 J^2+9
   S_{\rm BH}^4}{V_c}}}{6 \sqrt{3} \pi ^{5/2} c^2 S_{\rm BH}^4},
\end{equation} and
\begin{equation}
C_{P}=\frac{9 S_{\rm BH}^5 \sqrt[3]{\frac{P}{c^4}}-4 \pi ^4 J^2 S_{\rm BH} \sqrt[3]{c^2 P}}{12 \pi ^4 J^2
   \sqrt[3]{c^2 P}+9 S_{\rm BH}^4 \sqrt[3]{\frac{P}{c^4}}}.
\end{equation} Thus, for large temperatures \(T_{\rm H} \sim S_{\rm BH}\) and \(C_P \sim S_{\rm BH}\), i.e. these scalings are unchanged even after taking into account variations of \(G_d\). The corrections are therefore \(S_{\rm isoenthalpic-isobaric} = S_{\rm BH} - \frac{3}{2} \ln S_{\rm BH} + \cdots\), which are identical to those obtained in Eq. (\ref{extendedbtzcorr}) (for fixed \(G_d\)). \\

Now, considering the boundary, we recall that the gravity theory is dual to a \((1+1)\)-dimensional conformal field theory and the black hole entropy is reproduced from Cardy's formula:
\begin{equation}\label{cardy}
S_{\rm Cardy} = 2 \pi \sqrt{\frac{c\Delta}{6}} + 2 \pi \sqrt{\frac{\bar{c}\bar{\Delta}}{6}},
\end{equation} where \((c,\bar{c})\) are the central charges and \((\Delta,\bar{\Delta})\) are the conformal dimensions associated with the holomorphic and anti-holomorphic sectors, respectively \cite{BroHen}. The gravity parameters are related to CFT parameters as
\begin{equation}\label{Cg3}
c = \bar{c} = \frac{3l}{2G_3},
\end{equation}
\begin{equation}\label{Deltag3}
\Delta = \frac{(r_+ + r_-)^2}{16 G_3 l^2}, \hspace{5mm} \bar{\Delta} = \frac{(r_+ - r_-)^2}{16 G_3 l^2} ,
\end{equation} where \(r_+\) and \(r_-\) are the outer and inner horizon radii of the BTZ black hole. It is easy to verify that substituting Eqs. (\ref{Cg3}) and (\ref{Deltag3}) into Eq. (\ref{cardy}) gives \(S_{\rm Cardy} = \frac{\pi r_+}{2 G_3}\) which is just the Bekenstein-Hawking entropy \cite{BTZStrom}. Logarithmic corrections to the microcanonical entropy of the CFT were computed in \cite{Carlip} by considering logarithmic corrections to the Cardy's formula. The corrections are of the form suggested in Eq. (\ref{loggen}) with \(k = 3/2\), thereby matching with bulk. Thus, the coefficient \(k=3/2\) for the BTZ black hole in the fixed angular momentum ensemble seems quite robust, because it arises from various different frameworks.

\subsubsection{Black holes in AdS\(_5 \times S^5\)}\label{5dsec}
Consider charged black holes in AdS\(_5 \times S^5\) \cite{Chamblin}. In the high-temperature limit, the black hole mass scales with entropy as \(M \sim S_{\rm BH}^{4/3}\), meaning that \(T_{\rm H} \sim S_{\rm BH}^{1/3}\) and therefore, \(C_P \sim S_{\rm BH}\). These dependences are unaltered whether or not variations in \(G_d\) are considered. Thus, the logarithmic corrections to the isoenthalpic-isobaric entropy are \(S_{\rm isoenthalpic-isobaric} = S_{\rm BH} - \frac{5}{6} \ln S_{\rm BH} + \cdots\), i.e. \(k = 5/6\). \\

Next, let us consider the boundary. The dual theory corresponds to the \(\mathcal{N} = 4\) superconformal Yang-Mills theory at large \(N\), where \(N\) is the rank of the gauge group \(SU(N)\). The spacetime AdS\(_5 \times S^5\) can be thought of as being the near horizon limit of \(N\)-coincident D3-branes in type IIB supergravity (for large \(N\)) (see for example \cite{adscft}). In the decoupling limit, the theory is described by the massless excitations of open string states on the branes and the number of degrees of freedom, or the central charge scales with \(N^2\). These excitations are described by the \(\mathcal{N} = 4\) superconformal Yang-Mills theory at large \(N\) which contains \(8N^2\) bosonic and \(8N^2\) fermionic degrees of freedom\footnote{The particle content of the theory is \(N^2\) gauge fields, \(6N^2\) massless scalars, and \(4N^2\) Weyl fermions.}. Therefore, in this large \(N\) limit, the thermodynamics is described by the blackbody problem in \((3+1)\) dimensions, which are the world-volume dimensions of the D3-branes \cite{gubserkp} (see also \cite{KTentropy}), meaning that (up to constants) energy \(U \sim \mathcal{V} T_{\rm B}^4\) and entropy \(S \sim \mathcal{V} T_{\rm B}^3\), where \(\mathcal{V} = 2\pi^2 l^3\) and \(T_{\rm B}\) signifies blackbody temperature. This implies that the specific heat goes as
\begin{equation}
C_\mathcal{V} = T_{\rm B} \bigg(\frac{\partial S}{\partial T_B}\bigg)_\mathcal{V} \sim \mathcal{V} T_{\rm B}^3 \sim S,
\end{equation} and consequently, from Eq. (\ref{Smicrocan}), the logarithmic corrections to the microcanonical entropy due to energy fluctuations are given by \(S_{\rm mc} = S_0 - \frac{5}{6} \ln S_0 \cdots\), where \(S_0\) is the equilibrium value of \(S\), the field theory entropy. Thus, we reproduce the same coefficient \(k = 5/6\) as found in the bulk. \\

In these computations and those for the BTZ black hole, we have considered just one overall fluctuation coordinate, such as enthalpy \(H\) in the bulk, or energy \(U\) on the boundary. One may of course consider a larger number of fluctuation variables, including electric charge and central charge. 

\subsection{Fixed Newton's constant}
Let us consider a different route to holographic black hole chemistry, put up recently in \cite{vissernew}. In this approach, the Newton's constant is held fixed while cosmological constant is still allowed to vary. Thus, the bulk first law is just
\begin{equation}\label{bulknewsec}
\delta M = T_{\rm H} \delta S_{\rm BH} + V \delta P + \Phi \delta Q,
\end{equation} as discussed in section-(\ref{ExtSec}). This means, in the bulk, we continue to identify the black hole entropy with the isoenthalpic-isobaric entropy. Since we do not vary \(G_d\), the variations of bulk pressure are determined entirely in terms of variations of \(\Lambda\). Hence, \(V\) is the thermodynamic volume discussed in section-(\ref{ExtSec}) and differs from \(V_c\) introduced in the previous subsection. Now, on the boundary, the first law of thermodynamics has the same form as Eq. (\ref{firstlawboundhol}), i.e. 
\begin{equation}\label{boundnewsec}
\delta U = T \delta S - p \delta \mathcal{V} + \mu \delta c + \tilde{\Phi} \delta \tilde{Q},
\end{equation} where, as before \(p\) and \(\mathcal{V}\) are respectively the pressure and volume variables on the boundary, distinct from \(P\) and \(V\) of the bulk. Notice that the boundary first law admits work terms proportional to variations in both \(\mathcal{V}\) and \(c\)! This leads to a `degeneracy', in the sense that since both \(\mathcal{V}\) and \(c\) can be written in terms of \(l\), their variations are not truly independent and two distinct work terms are redundant. \\

The authors of \cite{vissernew} have proposed an interesting method to remove this redundancy which we briefly describe now. For \(d\) spacetime (bulk) dimensions, the CFT metric can be expressed as \(dl^2 = \omega^2(-dt^2 + l^2 d\Omega_{k,d-2}^2) \), where \(\omega\) is a parameter reflecting the conformal symmetry of the boundary theory. In the previous subsection, we implicitly considered the \(\omega = 1\) case. Notice that \(k = +1, 0 , -1\) specifies the topology of the \((d-2)\)-dimensional surface on which \(d\Omega_{k,d-2}^2\) is a line element. For instance, for \(k = +1\), we have \(d\Omega_{+1,d-2}^2\) describing the line element on a sphere, whereas \(k=0\) and \(k=-1\) indicate towards flat and hyperbolic topologies. Now, for a general scale parameter \(\omega\), the CFT volume scales as \(\mathcal{V} \sim (\omega l)^{d-2}\). This means, if apart from \(l\), we consider variations in \(\omega\), then the two work terms in the boundary first law can become independent due to the existence of an additional independent variable \(\omega\). To this end, we may define the thermodynamic parameters of the boundary theory to be related to those of the bulk as \cite{vissernew}
\begin{equation}\label{holmap}
S = S_{\rm BH}, \hspace{5mm} U = \frac{M}{\omega}, \hspace{5mm} T = \frac{T_{\rm H}}{\omega}, \hspace{5mm} \tilde{Q} = \frac{Ql}{\sqrt{G_d}}, \hspace{5mm} \tilde{\Phi} = \frac{\Phi \sqrt{G_d}}{\omega l},
\end{equation} wherein some straightforward manipulations shall reveal that Eqs. (\ref{bulknewsec}) and (\ref{boundnewsec}) can be mapped to each other. In a sense, this looks satisfying, because the bulk first law no longer contains a central charge term, which is rather a boundary quantity. Thus, in the bulk first law given in Eq. (\ref{bulknewsec}), there are three independent work terms corresponding to the three gravity parameters \(r_+\), \(l\) and \(Q\), while on the boundary [Eq. (\ref{boundnewsec})], there is an additional work term due to the existence of the additional parameter \(\omega\). Notice that \(\omega\) does not have any bulk analogue, and therefore its variations do not translate to the bulk. The following diagram summarizes the construction mapping the two first laws: 
\begin{center}
 \begin{tikzcd}[row sep=huge,column sep=huge]
 \delta H = T_{\rm H} \delta S_{\rm BH} + V \delta P + \Phi \delta Q \arrow[r, "S-{\rm function}"] \arrow[d, "{\rm Eqns~(\ref{holmap})~+~variations~of}~\omega"] 
& S_{\rm BH}(H,P,Q)\\ 
\delta U = T \delta S - p \delta \mathcal{V} + \mu \delta c + \tilde{\Phi} \delta \tilde{Q}   \arrow[r,  "S-{\rm function}"]
& S(U,\mathcal{V},c,\tilde{Q})   
\end{tikzcd}
\end{center}
Thus, while we identify the bulk entropy function to be that of the isoenthalpic-isobaric ensemble (equilibrium dictated by pressure), that of the boundary is the microcanonical ensemble (equilibrium dictated by volume). We consider the simple example of three-dimensional gravity. \\

For the BTZ black hole, the bulk computation involving the first law given in Eq. (\ref{bulknewsec}) gives \(k = 3/2\) as found in Eq. (\ref{extendedbtzcorr}) wherein, we have identified the black hole entropy with the isoenthalpic-isobaric entropy. Since on the boundary, the first law is given by Eq. (\ref{boundnewsec}), corrections to microcanonical entropy discussed previously, once again give \(k = 3/2\) leading to a match. We may consider higher-dimensional black holes too, such as the case considered in subsection-(\ref{5dsec}). Then, using Eq. (\ref{holmap}) and the techniques described so far, one gets the same numerical value for \(k\) leading to a match between the bulk and the boundary. 

\section{Conclusions}\label{Disc}
In this paper, we have discussed the computations of logarithmic corrections to the microcanonical entropy as well as isoenthalpic-isobaric entropy for various black holes in different ensembles. The approach adopted is the `one-loop approximation' in statistical mechanics, i.e. we consider the lowest order corrections to the density of states due to small fluctuations about equilibrium. While some new results were presented, we re-derived some older results for comparison. It was argued that the isoenthalpic-isobaric entropy has a more natural role in the context of black hole chemistry, where the cosmological constant leads to thermodynamic pressure. We found that the logarithmic corrections to the isoenthalpic-isobaric entropy in black hole chemistry agree with the corrections to the microcanonical entropy obtained in a framework where the cosmological constant is treated as a fixed parameter of the solution and is not interpreted as pressure. We also discussed microcanonical entropy corrections in black hole chemistry and defined a regularized microcanonical degeneracy for black holes saturating the reverse isoperimetric inequality. This leads to novel corrections to microcanonical entropy which cannot be obtained outside the framework of black hole chemistry \cite{open,CGQ}.\\

 In holographic black hole chemistry, irrespective of whether the Newton's constant is varied or not, one should compare corrections to the isoenthalpic-isobaric entropy in the bulk with those to the microcanonical entropy on the boundary. One aspect which has yet remained elusive is the correct matching of coefficients between bulk and boundary for terms which are of the type: \(\ln c\), where \(c\) is the number of degrees of freedom. As discussed in \cite{Carlip}, there are terms proportional to the logarithm of the central charge in corrections involving CFTs. We hope such issues will be reported on, in future.

\section*{Acknowledgements}
The author is grateful to Jasleen Kaur for carefully reading the manuscript, and to Chandrasekhar Bhamidipati and Sudipta Mukherji for several discussions related to this work.  The financial support received from the Ministry of Education (MoE), Government of India in the form of a Prime Minister's Research Fellowship (ID: 1200454) is gratefully acknowledged.

\appendix

\section{Computing \(D\) in two variables}\label{appA}
Let us explicitly compute \(D\) in a general setting, for an ensemble with two fluctuation coordinates. For dimensional consistency, we put \(x = \beta Y\) so that \(\beta Y\) is conjugate to \(X\). The entropy function here is defined as \(S(\beta,\beta Y) = \ln \mathcal{Z} + \beta E + \beta XY\), with the saddle-point equations:
\begin{eqnarray}
\bigg(\frac{\partial \ln \mathcal{Z} (\beta,\beta Y)}{\partial \beta}\bigg) &=& - E, \\
\bigg(\frac{\partial \ln \mathcal{Z} (\beta,\beta Y)}{\partial (\beta Y)}\bigg) &=& - X.
\end{eqnarray}
Thus, the covariance elements \(D_{11}\) and \(D_{12} (=D_{21})\) are
\begin{eqnarray}
 (\Delta E)^2  = \bigg(\frac{\partial^2 S(\beta,\beta Y)}{\partial \beta^2}\bigg)  &=& - \bigg(\frac{\partial E}{\partial \beta}\bigg)_{\beta Y}  , \label{EE2} \\
\langle (\Delta E)(\Delta X)\rangle = \bigg(\frac{\partial^2 S(\beta,\beta Y)}{\partial \beta \partial (\beta Y)}\bigg)  &=& - \bigg(\frac{\partial X}{\partial \beta}\bigg)_{\beta Y} . \label{EV2} 
\end{eqnarray}
It will be understood that the derivatives are evaluated at thermodynamic equilibrium, i.e. at \(\beta_0\) and \((\beta Y)_0\). In Eqs. (\ref{EE2}) and (\ref{EV2}), the derivatives with respect to \(\beta\) are taken at fixed \(\beta Y\) (rather than fixed \(Y\)). We may straightforwardly convert these derivatives to those at fixed \(\beta\) or \(Y\) as
\begin{eqnarray}
\frac{\partial E}{\partial \beta}\bigg|_{\beta Y} &=&  \frac{\partial E}{\partial \beta}\bigg|_{Y} - \frac{Y}{\beta} \frac{\partial E}{\partial Y}\bigg|_{\beta} ,
\end{eqnarray} and similarly for \(X\). Thus, Eqs. (\ref{EE2}) and (\ref{EV2}) become
\begin{eqnarray}
 (\Delta E)^2 &=& -  \frac{\partial E}{\partial \beta}\bigg|_{Y} + \frac{Y}{\beta} \frac{\partial E}{\partial Y}\bigg|_{\beta} ,\\
\langle (\Delta E)(\Delta X)\rangle &=&  -\frac{\partial X}{\partial \beta}\bigg|_{Y} + \frac{Y}{\beta} \frac{\partial X}{\partial Y}\bigg|_{\beta}. 
\end{eqnarray} Finally, \((\Delta X)^2\) (the element \(D_{22}\)) is given by
\begin{eqnarray}
(\Delta X)^2  = \bigg(\frac{\partial^2 S(\beta,\beta Y)}{\partial (\beta Y)^2}\bigg) = - \frac{\partial X}{\partial (\beta Y)}\bigg|_{\beta} = - \frac{1}{\beta} \frac{\partial X}{\partial Y}\bigg|_{\beta}. \label{VV3}   
\end{eqnarray} 
The covariances can be arranged to give the covariance determinant \(D =  (\Delta E)^2  (\Delta X)^2 - \langle (\Delta E)(\Delta X)\rangle^2 \). Using some general properties of partial derivatives and the Maxwell's relations, one can show after some calculation that 
\begin{equation}\label{D21}
D = -T^3 \bigg( \frac{\partial X}{\partial Y}\bigg)_T C_X,
\end{equation} where symbols have their usual meanings. Eq. (\ref{D21}) is equivalent to Eq. (\ref{D2}).

\section{Ensemble with \(Q^2\) as a thermodynamic variable}\label{Q2}
Since we are discussing various ensembles, it is interesting to remark on the situation where \(Q^2\), rather than \(Q\) is treated as a thermodynamic variable. In this case, the first law gets modified to \(\delta M = T_{\rm H} \delta S_{\rm BH} + \Theta \delta Q^2\) for a suitable variable \(\Theta\) which is defined as
\begin{equation}\label{Theta}
\Theta = \bigg( \frac{\partial M}{\partial Q^2} \bigg)_{S_{\rm BH}} = \frac{\sqrt{\pi} }{\sqrt{4 S_{\rm BH}}}. 
\end{equation}
The expression for the Hawking temperature remains unchanged. Notice that in this setting, the conjugate variable \(\Theta\) turns out to be independent of \(Q^2\). This means if one performs the Legendre transform \(\mathcal{E}(S_{\rm BH}, \Theta) = M(S_{\rm BH}, Q^2) - \Theta Q^2\), then \(\mathcal{E}\) becomes independent of \(\Theta\). One would expect that this new energy function satisfies the first law \(\delta \mathcal{E} = T_{\rm H} \delta S_{\rm BH} - Q^2 \delta \Theta\) but it does not, because
\begin{equation}
T_{\rm H} = \bigg( \frac{\partial \mathcal{E}}{\partial S_{\rm BH}} \bigg)_{\Theta} =  \frac{1}{\sqrt{16 \pi}} \Bigg[S_{\rm BH}^{-1/2} + \frac{3 S_{\rm BH}^{1/2}}{\pi l^2} \Bigg],
\end{equation} does not give the correct Hawking temperature [Eq. (\ref{chargednonextendedhawking})]. In fact, because one can't solve Eq. (\ref{Theta}) for \(Q^2\) means the Legendre transform is a singular one\footnote{One can however, resort to constraint analysis by treating Eq. (\ref{Theta}) as a constraint and impose it in the definition of \(\mathcal{E}\) by multiplying with a Lagrange multiplier. Then it is possible to reproduce the correct expression for Hawking temperature on-shell.} and the specific heat \(C_{\Theta} = 0\), because from Eq. (\ref{Theta}), fixing \(\Theta\) implies fixing \(S_{\rm BH}\) and vice versa. Furthermore, one gets \(\frac{\partial \Theta}{\partial Q^2} = 0\) meaning that the \(Q^2\)-susceptibility is undefined. Thus, one cannot obtain logarithmic corrections due to fluctuations of \(Q^2\) at least without any regularization. On the other hand, if one considers energy fluctuations alone, then in the high-temperature limit one gets logarithmic corrections identical to Eq. (\ref{abcd}).

\end{document}